\documentclass[aps,prb,twocolumn,showpacs,amsmath,amssymb]{revtex4}
\usepackage{graphicx}
\usepackage{dcolumn}
\usepackage{bm}
\usepackage{color}

\begin{document}
\title{Relevance of complete Coulomb interaction matrix for the Kondo problem: \\ Co impurity on Cu(111)}
\author{E. Gorelov$^{1}$, T. O. Wehling$^{2}$,  A. N. Rubtsov$^{3}$,  M. I. Katsnelson$^{4}$, and A. I.
Lichtenstein$^{2}$}
\affiliation{
 $^{1}$ Institut f\"ur
Festk\"orperforschung and Institute for Advanced Simulation,
Forschungzentrum J\"ulich, 52425 J\"ulich,
Germany\\
$^{2}$ I. Institute of Theoretical Physics, University of Hamburg,
20355 Hamburg, Germany\\
$^{3}$ Department of Physics, Moscow State University, 119992
Moscow, Russia\\
$^{4}$Institute for Molecules and Materials, Radboud University of
Nijmegen, 6525 ED Nijmegen, The Netherlands }
\date{\today}

\begin{abstract}
The electronic structure of a prototype
Kondo system, a cobalt
impurity in a copper host is calculated with accurate taking into
account of correlation effects on the Co atom. Using the recently developed
continuous-time QMC technique, it is possible
to describe the Kondo resonance with a complete four-index Coulomb
interaction matrix. This opens a way for completely
first-principle calculations of the Kondo temperature. We have
demonstrated that a standard practice of using a truncated Hubbard
Hamiltonian to consider the Kondo physics can be quantitatively
inadequate.
\end{abstract}
\maketitle

\section*{Introduction}

Scanning tunneling microscopy (STM) has become one of the most
basic tools for the manipulation of matter at the atomic scale.
Although this experimental technique has reached maturity, the
detailed theoretical understanding of experimental data is still
incomplete and/or contradictory. One of the most famous examples
of atomic manipulation is associated with the surface Kondo effect
observed when transition metal ions (like Co) are placed on a
metallic surface (such as Cu $(111)$) \cite{Madhavan98,jli_98}.
The surface Kondo effect is the basis for the observation of
surprising phenomena like quantum mirages \cite{coral}, and has
attracted a lot of attention and interest in the last few years.
Early interpretations of these observations were based on the
assumption that only surface states of Cu $(111)$ are involved in
the scattering of electron waves by the Co adatoms
\cite{Fiete03,Agam00,Porras01}. However, later experiments with Co
atoms on the Cu (100) surface (that does not have any surface
state) \cite{Knorr02}, or in Cu (111) but close to atomic surface
steps (that affect the surface states) \cite{LimotCM} have
indicated that bulk rather than surface states are responsible for
the Kondo effect in these situations. The latter can be important
for fine tuning of surface electronic structure, with potential
applications to nanotechnology. A recent study of CoCu$_n$
clusters on Cu (111) demonstrated this tunablilty by atomic
manipulation and showed that each atom in the vicinity of the magnetic
impurity matters for determining the Kondo effect \cite{Neel}.
Moreover, the relevance of the Kondo effect for the electronic
structure of metal surfaces themselves was demonstrated by the
discovery of a sharp density of states peak on the Cr $(001)$
surface and its possible interpretation as an orbital Kondo
resonance \cite{Cr,Cr1,Cr2}.

At the same time, when calculating the Kondo temperatures for real
electronic structures a mapping onto one-orbital Anderson impurity
model \cite{Jones06} was used. The realistic atomic geometry of Kondo
systems plays a crucial role in complex electronic properies
\cite{Neel,Vitali08} and it is, a priori, not obvious that
a one-orbital Anderson impurity approach is
sufficient: even the two-orbital Anderson model demonstrates
Kondo physics essentially different from the single-orbital one
\cite{Zhuravlev}. A recent theoretical investigation of Fe
impurities in gold and silver showed that the proper Kondo model
corresponds to a S=3/2 spin state\cite{Costi09}. A realistic,
multi-band consideration of correlation effects in specific solids
is possible in the framework of the Local Density Approximation +
Dynamical
Mean-Field Theory (LDA+DMFT) approach (for review, see Ref. %
\onlinecite{KotSavr06}). However, formally accurate Quantum Monte Carlo
(QMC) calculations are always done with taking into account only the diagonal
part of Coulomb interaction \cite{KatLi00,LLSr2RuO4}, even with realistic
hybridization functions obtained in the LDA. This approximation is, strictly
speaking, uncontrollable. At the same time, approximate schemes working with
the complete Coulomb interaction matrix, such as the perturbative scheme
\cite{SPTF} which is frequently used to calculate electronic structure of
transition metals and alloys \cite{KKR,DiMarco} are not sufficient to
reproduce so subtle correlation features like the Kondo effect, properly. As
for the exact diagonalization \cite{Cr,Cr1} or numerical renormalization
group \cite{Cr1,Zhuravlev,Bulla} methods they are hardly applicable, due to
computational problems, for more than two orbitals per impurity.

The recent progress in continuous time QMC scheme (CT-QMC)
\cite{ctqmc,hybrqmc} makes it perspective to treat the complicated
Kondo systems \cite{triangle}. Here we will apply this method to
calculate Kondo temperatures as well as spectral functions for the
case of a Co impurity in bulk Cu, in a Cu $(111)$ surface and on top
of a Cu $(111)$ surface. In contrast with all previous calculations
we will work with an accurate complete Coulomb interaction
$U$-matrix for correlated $d$ orbitals.
The latter can be calculated from first principles in
a parameter-free way by the GW technique \cite{GW} so this
approach is completely \textit{ab initio}. Moreover, the CT-QMC
method allows to work, without any essential difficulties, even
with the rigorous frequency-dependent $U$-matrix. As the first
step, we present calculations for the static $U$-matrix, but this
restriction is purely technical and can be relatively easily
removed in the future, with a growth of available computer
resources.

\section{Multi-orbital CT-QMC formalism}

The multi-orbital impurity problem with a general $U$-matrix is described by
the effective action
\begin{eqnarray}  \label{Sinit}
S_{imp}=&&S_{0}+S_{int}  \nonumber \\
=&&-\sum_{ij\sigma }\int_{0}^{\beta }\int_{0}^{\beta }{\mathcal{G}}%
_{ij}^{-1}(\tau -\tau ^{\prime })c_{i\sigma }^{\dagger }c_{j\sigma
}d\tau
d\tau ^{\prime }  \nonumber \\
&&+\frac{1}{2}\sum_{ijkl\sigma \sigma ^{\prime }}\int_{0}^{\beta
}U_{ijkl}c_{i\sigma }^{\dagger }c_{j\sigma ^{\prime }}^{\dagger
}c_{k\sigma ^{\prime }}c_{l\sigma }d\tau
\end{eqnarray}
where $i,j,k,l$ are orbital indices, and $\sigma ,\sigma ^{\prime }$ are spin
indices, ${\mathcal{G}}_{ij}$ is the local non-interacting Green function
for correlated orbitals obtained from the Density Functional Theory (DFT)
with the help of optimal projection operator to the impurity d-states:
\begin{equation}
 {\mathcal{G}}_{ij}(i\omega _{n})=\sum_{n\mathbf{k}}\frac{\left\langle
d_{i}|\psi _{n\mathbf{k}}\right\rangle \left\langle \psi _{n\mathbf{k}%
}|d_{j}\right\rangle }{i\omega _{n}+\mu -\varepsilon _{n\mathbf{k}}}; \label{eqn:G0}
\end{equation}
here $\varepsilon _{n\mathbf{k}}$ is the energy spectrum and $\psi _{n%
\mathbf{k}}$ is the corresponding wave function of our system (metal host
with magnetic impurity), described by $d_{i}$ localized orbitals, and $%
U_{ijkl}$ is the Coulomb interaction matrix element:
\begin{equation}
U_{ijkl}=\left\langle i_{1}j_{2}\left| V_{12}^{ee}\right|
k_{2}l_{1}\right\rangle   \label{Umat}
\end{equation}
here $i_{1}\equiv d_{i}\left( \mathbf{r}_{1}\right) $ is local orthogonal
wave function for correlated orbitals and\ $V_{12}^{ee}$ is screened
spin-independent Coulomb interaction between electrons at the coordinates $%
\mathbf{r}_{1}$ and $\mathbf{r}_{2}$. We used standard quasiatomic
LDA+U parametrization of Coulomb matrix for d-electron via
effective Slater integarls or average Coulomb parameter U and
exchange parameter J as described in Ref.\onlinecite{LDAU}. We
choose the orbital basis related to spherical harmonics to be sure
that magnetic orbital
quantum numbers in $U_{ijkl}$ matrix are satisfied the following sum rule: $%
i+j=k+l$. In this case we will get rid of so-called three-site terms like $%
U_{ikkl}$ with $i\neq l$ which turns out to result in a strong
sign problem in QMC calculations with real spherical harmonics.

Following the general CT-QMC scheme \cite{ctqmc} we expand the partition
function around the Gaussian part of our multiorbital action Eq.(\ref{Sinit})
which gives the fermionic determinant over the non-interacting Green
functions with the rank $2n$:

\begin{widetext}

\begin{equation}
\frac{Z}{Z_{0}}=\sum_{n}\frac{(-1)^{n}}{n!2^{n}}\sum_{\{ijkl\sigma \sigma
^{\prime }\}}\int_{0}^{\beta }d\tau _{1}...\int_{0}^{\beta }d\tau
_{n}U_{i_{1}j_{1}k_{1}l_{1}}...U_{i_{n}j_{n}k_{n}l_{n}}\det {\mathcal{G}}%
^{2n\times 2n}  \label{Zexp}
\end{equation}

\end{widetext}

In order to minimize the number of different interaction vertices we group
different matrix elements of the multiorbital Coulomb interactions which
have a similar structure of fermionic operators. Since the $U_{ijkl}$ matrix
elements are spin independent,
one should look over all possible combinations of
orbital and spin indices,
to generate all terms for the interaction in
the action Eq.(\ref{Sinit}).
Some combinations can violate the Pauli principle
and should be removed. For CT-QMC algorithm it is useful to
represent the interaction Hamiltonian in the following form: $%
U_{ijkl}c_{i\sigma }^{\dagger }c_{l\sigma }c_{j\sigma ^{\prime }}^{\dagger
}c_{k\sigma ^{\prime }}$.

The interaction terms can be transformed to the desired form, depending on
relations between spin and orbital indices:

(i) if $\sigma \neq \sigma ^{\prime }$, we can just commute $c_{l\sigma }$
and $c_{k\sigma ^{\prime }}$ and then $c_{l\sigma }$ and $c_{j\sigma
^{\prime }}^{\dagger }$. Another combination of indices, that allows the
same commutation, is the following: $\sigma =\sigma ^{\prime }$, $i\neq j$
and $k\neq l$ (the latter two are following from the Pauli principle), and also $%
j\neq l$. These terms we can transform to the following desirable
representation:

\begin{equation}
H_{ijkl}^{int1}=U_{ijkl}c_{i\sigma }^{\dagger }c_{l\sigma
}c_{j\sigma ^{\prime }}^{\dagger }c_{k\sigma ^{\prime }}.
\label{sym1}
\end{equation}
\ \

(ii) in the case when $\sigma =\sigma ^{\prime }$ and $j=l$ \ we can commute
$c_{k\sigma ^{\prime }}$ and $c_{j\sigma ^{\prime }}^{\dagger }$, since in
this case $i\neq j$ and $k\neq l$ due to the Pauli principle:

\begin{equation}
H_{ijkl}^{int2}=-U_{ijkl}c_{i\sigma }^{\dagger }c_{k\sigma
}c_{j\sigma }^{\dagger }c_{l\sigma }  \label{sym2}
\end{equation}

After generating all this terms it is
useful to collect and \textit{symmetrize} all the terms with identical
and equivalent (i.e. $U_{ijkl}c_{i\sigma
}^{\dagger }c_{j\sigma }c_{k\sigma ^{\prime }}^{\dagger }c_{l\sigma ^{\prime
}}$ and $U_{klij}c_{k\sigma ^{\prime }}^{\dagger }c_{l\sigma ^{\prime
}}c_{i\sigma }^{\dagger }c_{j\sigma }$)
quantum numbers.
\newline


In order to reduce the fermionic sign problem we introduce additional
parameters, $\alpha $, to optimize the splitting of the Gaussian and
interaction parts of the action Eq.(\ref{Sinit})

\begin{widetext}
\begin{eqnarray}
&&S_{0}=\sum_{ij\sigma }\int_{0}^{\beta }\int_{0}^{\beta }\left( -{\mathcal{G%
}}_{ij}^{-1}(\tau -\tau ^{\prime })+\frac{1}{2}\sum_{\{kl\sigma ^{\prime
}\}}\alpha _{kl}^{\sigma ^{\prime }}(U_{ilkj}+U_{lijk})\delta _{\tau \tau
^{\prime }}\right) c_{i\sigma }^{\dagger }c_{j\sigma }d\tau d\tau ^{\prime
}~,  \label{S0m} \\
&&S_{int}=\frac{1}{2}\sum_{\{ijkl\sigma \sigma ^{\prime }\}}\int_{0}^{\beta
}U_{ijkl}(c_{i\sigma }^{\dagger }c_{l\sigma }-\alpha _{il}^{\sigma
})(c_{j\sigma ^{\prime }}^{\dagger }c_{k\sigma ^{\prime }}-\alpha
_{jk}^{\sigma ^{\prime }})d\tau .  \nonumber
\end{eqnarray}
\end{widetext}

One can see, that the first item in (\ref{S0m}) on Matsubara
frequences corresponds to bare Green's function
\begin{equation}
{\mathcal{G}}_{ij}^{-1}=\left( i\omega _{n}+\mu \right) \delta _{ij}-\Delta
_{ij}(\omega _{n})  \label{G00}
\end{equation}
where $\Delta $ is the hybridization matrix. The second term is just a
constant which we can absorb to the new chemical potential $\tilde{\mu}$.
Therefore we can rewrite the bare Green function in the following matrix
form:
\begin{equation}
\tilde{\mathcal{G}}^{-1}=\left( i\omega _{n}+\tilde{\mu}\right) \mathbf{1}-%
\mathbf{\Delta },  \label{G0}
\end{equation}
The optimal choice of parameters $\alpha _{ij}^{\sigma }$ would lead to
effective reduction of interaction terms in the action Eq. (\ref{S0m}) and
therefore minimization of average perturbation order in Eq. (\ref{Zexp}).

Note that relation between $\tilde{\mathcal{G}}$ and ${\mathcal{G}}$ can be
represented from Eq. (\ref{S0m}) in the following spin and orbital matrix
form:
\begin{equation}
\tilde{\mathcal{G}}^{-1}={\mathcal{G}}^{-1}-\left\langle \widehat{\alpha }%
\hat{U}\right\rangle .  \label{G0m}
\end{equation}
Here we used the fact that $U_{ilkj}=U_{lijk}$ following from the definition
of the Coulomb matrix elements (\ref{Umat}).

We also need to minimize the fermionic sign problem which finally leads us to
such expression for diagonal alpha parameters
\begin{equation}
\alpha _{\sigma }^{ii}+\alpha _{\sigma ^{\prime }}^{jj}=\bar{\alpha},
\label{Alph_diag}
\end{equation}
corresponding to to the following interaction fields $U_{ijji}n_{i\sigma }n_{j\sigma ^{\prime
}}$. The
$\bar{\alpha}$ has to be found iteratively in order to get a proper
occupation number of correlated electrons. In the case of half-filled
one-band Hubbard model $\bar{\alpha}=1$ leads to the correct chemical
potential shift of the $\frac{U}{2}$ and average $\alpha =\frac{1}{2}$ which
corresponds to the Hartree-Fock substraction.
For non-diagonal alpha's which correspond to the fields of general
form $U_{ijkl}c_{i\sigma }^{\dagger }c_{l\sigma }c_{j\sigma ^{\prime
}}^{\dagger }c_{k\sigma ^{\prime }}$, where $i\neq l$ and $j\neq k$ we find
the following optimal condition:
\begin{equation}
\alpha _{\sigma }^{ij}+\alpha _{\sigma ^{\prime }}^{kl}=0  \label{Alph_nd}
\end{equation}

Since we symmetrize the interaction $U$ matrix it is necessary to
extend the definition of the ${\hat{\alpha}}$ \ matrix to
keep all the terms in the interaction part of initial action ( the
last item in Eq. (\ref{S0m})). It can be done in the following way
\cite{ctqmc,Assaad07}: for every $U_{ijkl}$ field in $50\%$ of
updates we deliver the $\alpha $ parameters as \newline $\alpha
^{il}=\alpha _{diag}$, $\alpha ^{jk}=\bar{\alpha}-\alpha _{diag}$,
and in another $50\%$ as
$\alpha ^{il}=\bar{\alpha}-\alpha _{diag}$, $\alpha ^{jk}=\alpha _{diag}$
for the case of $i=l$ and $j=k$. For non-diagonal fields, i.e. $i\neq l$ and
$j\neq k$
$\alpha ^{il}=\alpha _{nd}$, $\alpha ^{jk}=-\alpha _{nd}$,
with $50\%$ probablility and
$\alpha ^{il}=-\alpha _{nd}$, $\alpha ^{jk}=\alpha _{nd}$
otherwise. It was found that the sign problem is eliminated in the case when
$\alpha _{diag}<0$ and ${\bar{\alpha}}\geq 1$ for occupancy $n\geq \frac{1}{2%
}$ per
 state
and $\alpha _{diag}>0$, ${\bar{\alpha}}<1$ otherwise. The
optimal choice of $\left| \alpha _{diag}\right| $ parameter is few percent
of $\left| \bar{\alpha}\right| $ to keep minimal average perturbation order.
Another problem is a proper choice of non-diagonal $\alpha _{nd}$ parameter.
It is easy to see that $\alpha _{nd}$ is proportional to acceptance
probability of non-diagonal field in the case where
corresponding bare Green function
$%
{\mathcal{G}}_{jk}=0$. Since these processes are unphysical, the
natural choice is $\alpha _{nd}=0$. But it leads to division
by zero in the updating the inverse Green function matrix
\cite{ctqmc}. On the other hand increasing the $\alpha _{nd}$
parameter causes a sign problem. We find a reasonable choice of
$\alpha _{nd}$ to be on the order of $10^{-4}$. Moreover for some
special cases like the atomic limit, where
${\mathcal{G}}_{mm}(\tau )$
is constant,
a small noise should be added to
all the $\alpha $ parameters to avoid numerical divergency.

\section{Results}

The Co-Cu system is treated as five-orbital impurity model
representing $3d$ electronic shell of the cobalt atom hybridized
with a bath of a conduction Cu electrons. The bath Green function
was obtained using the first-principle density-functional theory
within the supercell approach. For Co impurity atoms in the bulk
as well as in / on the Cu (111) surface the bath Green functions
were obtained using the Vienna-Ab-Initio simulation package
(VASP)\cite{gkr_94,gkr_99} using the projector augmented wave
(PAW) basis sets\cite{pbl_94}. The density functional calculation
for cobalt impurity in the bulk was carried out using CoCu$_{63}$
supercell structure with lattice constant corresponding to
pure copper. The surfaces were modelled by supercells of Cu (111)
slabs containing 5 Cu layers with $2\times 2$ and $3\times 4$
lateral extension for Co in and on the surface, respectively. The
PAW basis naturally provides the projectors $\left\langle
d_{i}|\psi _{n\mathbf{k}}\right\rangle$ required in Eq.
(\ref{eqn:G0}). In using these PAW projectors, directly, we employ
here the same representation of localized orbitals as used within
the LDA+U-scheme implemented in the VASP-code itself or as
discussed in the context of LDA+DMFT in Ref.
\onlinecite{PAW-DMFT}.

For the problem of a single Co impurity in a bulk copper matrix the basis set of
spherical harmonics $Y_{lm}$ is used. In this basis the interaction part of
the hamiltonian contains only terms of the following form: 
diagonal density-density like $H_{int}^{diag}=U_{ijji}n_{i\sigma }n_{j\sigma
^{\prime }}$ , where $n_{i\sigma }=c_{i\sigma }^{\dagger }c_{i\sigma }$ and
non-diagonal
$H_{int}^{nd}=U_{ijkl}c_{i\sigma }^{\dagger }c_{j\sigma
^{\prime }}c_{k\sigma ^{\prime }}^{\dagger }c_{l\sigma },$ where $i\neq j$
and $k\neq l$.
The Coulomb matrix for the $d$-electron shell in the basis of complex
harmonics contains 45 non-equivalent diagonal terms.
Non-diagonal terms can be further classified to a spin-flips, where $i=l$, $%
j=k$ and the most general four-orbitals interactions, where this condition
is not fulfilled. Notice, that pair-hopping terms ($i=k$, $j=l$) are
restricted by symmetry in this basis. In 
description of $d$-electron shell 
we have to involve 20 non-equivalent spin-flips and 64 terms of the most
general form.

To find the effects, caused by non-diagonal terms, 
we used two different interaction Hamiltonian.
First,
interaction with only diagonal terms 
was used. In this case there is no sign problem.
Then, the complete Coulomb interaction matrix of the $3d$-electron shell of the
cobalt atom with 129 terms was included.\newline

\begin{figure}[tbp]
\begin{center}
\includegraphics[width=0.5\textwidth]{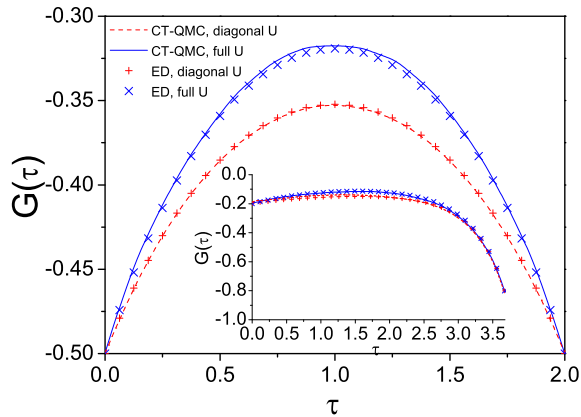}
\end{center}
\caption{(color online) Comparison with ED in the atomic limit (without
hybridisation to the bath of free electrons). Main graph: $U=1$
eV, $J=0.4$ eV $\beta =2$ eV$^{-1}$, for 5-orbital impurity at
half-filling; inset: $U=2$ eV, $J=0.7$ eV, $\beta =3.7$ eV$^{-1}$,
\ for 5-orbital impurity with 8 electrons.} \label{TBvsED}
\end{figure}

As a benchmark we use impurity problem in the atomic limit,
since it can be compared with the result of exact diagonalization (ED) method.
The results imaginary time Green function for 5-orbitals
model with different chemical potentials corresponding to the $d^{5}$
and $d^{8}$ configurations are shown
in the Fig. \ref{TBvsED} in comparison with ED results.
The significant difference between density-density
(diagonal) 
interaction and the full vertex 
can be found both at half-field case with relatively high
temperature with the $U=1$ $eV$, $J=0.4$ $eV$, $\beta=2$ $eV^{-1}$
and at non-symmetric case even for lower temperature. Note that in
the $d^{8}$ and $d^{7}$ cases the many-body ground-state have
different symmetry for diagonal interactions
and non-diagonal full vertex. 
The results for $d^{8}$ configuration with the interaction parameters $U=2$ $%
eV$, $J=0.7$ $eV$, $\beta=3.7$ $eV^{-1}$ are shown
in the insert to the Fig. \ref{TBvsED}.
The difference between Green function of the interacting system
with full Coulomb interaction and density-density one is visible on the $%
G(\tau)$. We find a very good agreement between CT-QMC results and ED
solution.

\begin{figure}[tbp]
\begin{center}
\includegraphics[width=0.5\textwidth]{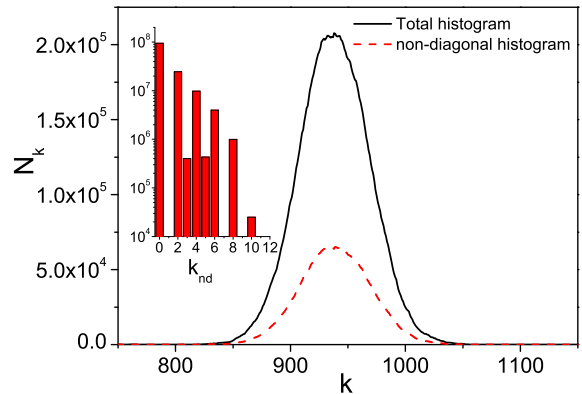}
\end{center}
\caption{(color online) Histograms of Monte-Carlo distributions for average
perturbation order. Main graph: $U=4$ eV, $J=0.7$ eV, $\beta=10$
eV$^{-1}$ for 5-orbital impurity coupled to realistic Cu-bath with
7 electrons; inset: $U=4$ eV, $J=0.7$ eV, $beta =1$ eV$^{-1}$ in
the case of 5-orbital impurity model, coupled to semi-elliptical
bath with bandwidth $W=0.5$ eV at the half-filling} \label{hist}
\end{figure}

In the inset of Fig. \ref{hist} we show the distribution of
non-diagonal terms, i.e. the contribution of Coulomb fields of the
form (\ref {sym1}) to the resulting Green function. The zero entry
of this histogram counts the number of steps when all the fields
contributing to the fermionic determinant (\ref{Zexp}) were of
density-density type. The entry with index $2$ show us the number
of steps where the average (\ref{Zexp}) was containing two
spin-flip type fields (\ref{sym1}). Such situation takes
place, for example, when one Coulomb field representing spin-flip $%
c_{i\uparrow }^{\dagger }c_{j\downarrow }c_{i\downarrow }^{\dagger
}c_{j\uparrow }$ process was used to construct the determinant.

One can see in the insert of the Fig. \ref{hist}, that only even orders of
interaction histogram have large acceptance probability at high temperature
and even the tenth order in non-diagonal interactions has non-zero contribution.
The 3-rd and 5-th order contributions exists due to the finite $\alpha^{nd}$ parameter.

\

Typical distribution of the perturbation order (5-orbital AIM with
$7$ electrons, $U=4$ $eV$, $J=0.7$ $eV$, $\beta=10$ $eV^{-1}$) is
shown in
Fig. \ref{hist}, main plot. Dash line denotes the
perturbation order during accepted steps that involved
non-diagonal fields. The coincidence of distributions maxima of
both histograms demonstrate that the acceptance rate mostly
depends on diagonal interactions.

For many-body calculations of the Co impurity in the Cu matrix we need to find the
effective $d$-orbital chemical potential which defines the number of
3d-electrons of cobalt. The particular electronic configuration of a Co atom
in a copper matrix is unknown, but the DFT results ($n_{d}=7.3$) give us an
evidence that it is close to $d^{7}$ configuration. Therefore we performed
all impurity calculations for cobalt  $d^{7}$ configuration.

\begin{figure}[tbp]
\begin{center}
\includegraphics[width=0.5\textwidth]{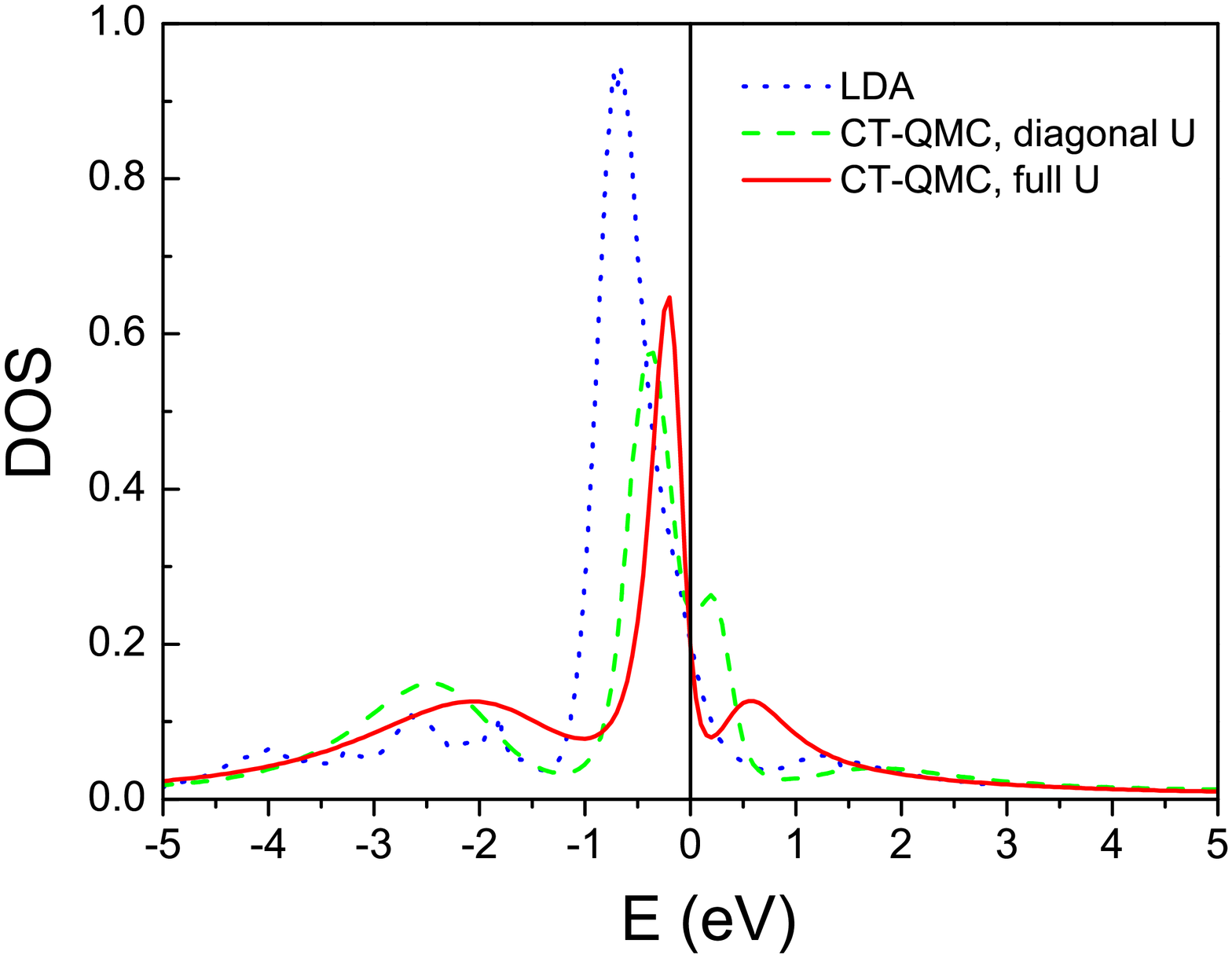}
\end{center}
\caption{(color online) Total DOS of 3d orbital of Co atom embedded in Cu matrix.
Model parameters: $U=4$ eV, $J=0.7$ eV, $\beta =10$ eV$^{-1}$ for
5-orbital impurity with 7 electrons.} \label{doses5b}
\end{figure}

The results of the CT-QMC calculations for $U=4$ eV and $J=0.7$ eV are
presented
in  Fig.{\ref{doses5b}}
compared to the bare
impurity density of states for cobalt impurity in the bulk. There
is a pronounced difference between Kondo-like resonance near the
Fermi level.In the case of full $U$-vertex
it becomes more narrow and
located much closer to the Fermi level. The sign problem for
realistic five-band model depends crucially on the symmetry of
coulomb Vertex $U_{ijkl}$ and
magnitude of non-diagonal terms
in the bath Green functions ${\mathcal{G}}_{ij}$. The most serious
problem is related with non-diagonal terms of U-matrix, therefore
we use a basis of complex spherical harmonics. In this case there
is no so-called three-cite terms or correlated hopping, e.g.
$U_{ikkl}$.
On the other hand, in this basis, the bath Green-function matrix
${\mathcal{G}}_{ij}$ for d-electrons has two non-diagonal elements
in the bulk of cubic crystals and much more on the surface and in
the first layer. Moreover there are lot of small four-site terms
$U_{ijkl}$ which result in a large sign problem for surface-adatom
calculations. The sign problem for a Co impurity in the
bulk is not large and average sign is between 0.90 and 0.97
depends on the simulation temperature.

In the case of non-diagonal interaction we used so-called cluster
steps which correspond to complex
Monte-Carlo updates with more
than one additional interaction field. This scheme became
essential for spin-flip like interaction or more general U-vertex
which can contribute to the Green-function only in the second or
higher order ''diagammatic'' expansion and this can let the
Monte-Carlo process to explore all the phase space. We note that
probability of non-diagonal terms drastically decrease with
increasing the hybridization to the bath.
Nevertheless, at least
for three-band benchmarks we found remarkable effect
of the spin-flip terms if the bath Green function has
peaks in the vicinity of the Fermi level on the distance of the
order of $J$.

\begin{figure}[tbp]
\begin{center}
\includegraphics[width=0.5\textwidth]{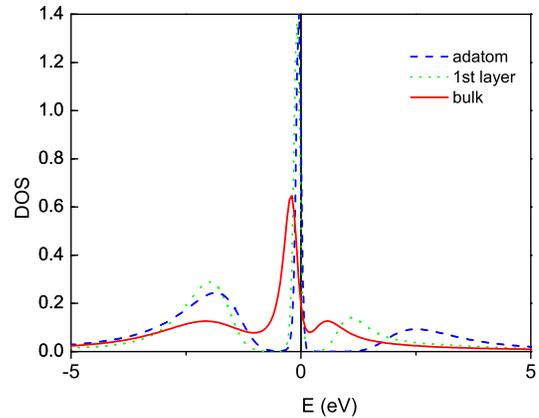}
\end{center}
\caption{(color online)Total DOS of 3d orbital of Co atom embedded in the bulk
of Cu, into 1-st layer and Co-adatom on the Cu(111) surface. Model
parameters: $U =4$ eV, $J =0.7$ eV, $\protect\beta =10$ eV$^{-1}$ for
5-orbital impurity with 7 electrons.} \label{3DOS}
\end{figure}
\bigskip

We estimated the renormalization factor $Z= (1-d\Sigma/dE)^{-1}$
for $U=4.5$ eV, $J=0.7$ eV and $\beta =10$ eV$^{-1}$ and find
$Z_{t_{2g}}=0.5$ and $Z_{e_{g}}=0.4$ which
shows the reasonable strong interaction of Co d-electrons. We estimate the
Kondo temperature (T$_{K}$) using the temperature dependence of FWHM for
resonance near Fermi level. Since our simulation temperature is very high
compared to T$_{K}$ we can get only order of magnitude of T$_{K}=0.1$ eV,
which is reasonable for Co-impurity systems.

We also performed the CT-QMC calculation of cobalt impurity on the surface
of Cu(111) and embedded into the first copper layer. In contrast to the bulk
system the surface one has a large sign problem, related with the
relativelly large non-diagonal elements of the bath Green functions.
Although changing of the sign is a very rare event (less then 0.03\% of the
accepted steps) and we used a simple constrained sign calculations.
Comparison of the different spectral functions for the bulk, surface and
fist-layer cobalt impurity is presented in Fig. \ref{3DOS}. One can see
clearly the change of the Kondo resonance width as a function of reduced
dimensionality.

\section*{CONCLUSIONS}

In conclusion, we perform the continuous time QMC calculation of
realistic 5-orbital Co impurity in copper and discuss the
relevance of non-diagonal part of Coulomb matrix in the Kondo
problem. Comparing Figs. \ref{doses5b} and \ref{3DOS} we find that
non-density-density terms in the Coulomb vertex are required to
obtain quantitative predictions of spectral functions and related
properties. The position of the Hubbard peaks and the Kondo peak
is markedly changed by
spin-flips and other non-diagonal terms of the Coulomb vertex. Thus, obtaining
sensitive observables like Kondo temperatures quantitatively
requires accounting for these terms. On the other hand
hybridization effects like bringing the Co impurity from bulk to
the surface and having it on top of the surface can be quite
drastic: As Fig \ref{3DOS} shows, the sharpening of the Kondo
resonance and the shifting of the Hubbard bands is much stronger
when going from bulk to the surface then on switching on the
non-diagonal part of Coulomb matrix. Only the qualitative
overall shape the DOS and its response to strong hybridization changes are
well described by density-density type terms of the Coulomb
vertex.

\section*{ACKNOWLEDGEMENTS}
The authors acknowledge a financial support from DFG SFB-668 (Germany)
RFFI (Russia) and FOM (The Netherlands).

\end{document}